\documentclass[letter,traditabstract]{aa} % for the letters 
\usepackage{graphicx}
\usepackage{txfonts}
\usepackage{natbib}
\usepackage[usenames,dvipsnames]{color}
\usepackage{epstopdf}
\usepackage{verbatim}
\usepackage{lscape}

\def\lsim{\mathrel{\rlap{\lower4pt\hbox{$\sim$}}
    \raise1pt\hbox{$<$}}}                % less than or approx. symbol
\def\gsim{\mathrel{\rlap{\lower4pt\hbox{$\sim$}}
    \raise1pt\hbox{$>$}}}                % greater than or approx. symbol

\def\pipe{Pipe Nebula}

\def\Av{\mbox{$A_{\rm V}$}}

\def\kms{\mbox{km~s$^{-1}$}}

\def\vlsr{${\tt v}_{\rm LSR}$}
\def\asec{.$\!''$}

\def\cdh{C$_2$H}
\def\cthd{c--C$_3$H$_2$}
\def\tcthd{c--H$^{13}$CCCH}
\def\cqh{C$_4$H}
\def\chtcdh{CH$_3$C$_2$H}
\def\cdo{C$^{18}$O}
\def\tcdo{$^{13}$C$^{18}$O}
\def\hco{HCO}
\def\hcom{HCO$^+$}
\def\htcom{H$^{13}$CO$^+$}
\def\hcdom{HC$^{18}$O$^+$}
\def\cs{CS}
\def\hdcs{H$_2$CS}

\def\hcsp{HCS$^+$}
\def\so{SO}
\def\tso{$^{34}$SO}

\def\sod{SO$_2$}
\def\ocs{OCS}
\def\hcn{HCN}
\def\hnc{HNC}
\def\htcn{H$^{13}$CN}
\def\hcqn{HC$^{15}$N}
\def\hntc{HN$^{13}$C}
\def\hqnc{H$^{15}$NC}
\def\hctn{HC$_3$N}
\def\hnco{HNCO}
\def\cthdt{C$_3$HD}
\def\nhdd{NH$_2$D}
\def\dctn{DC$_3$N}
\def\hocom{HOCO$^+$}
\def\hccnc{HCCNC}

\begin{document}
   \title{Chemical Differentiation toward the Pipe Nebula Starless Cores}

   \author{
P.\ Frau\inst{1}
\and
J.\ M.\ Girart\inst{1}
\and
M.\ T.\ Beltr\'an\inst{2}
}

\institute{
Institut de Ci\`encies de l'Espai (CSIC-IEEC), Campus UAB, 
Facultat de Ci\`encies, Torre C5p, 08193 Bellaterra, Catalunya, Spain\\
\email{[frau;girart]@ice.cat}
\and
INAF-Osservatorio Astrofisico di Arcetri, Largo Enrico Fermi 5, 50125 Firenze, Italy\\
\email{mbeltran@arcetri.astro.it}
}

\date{Received 8 December 2011 / Accepted 22 December 2011}

\abstract{We used the new IRAM~30-m FTS backend to perform an unbiased $\sim$15~GHz wide
survey at 3~mm toward the \pipe\ young diffuse starless cores. We found an unexpectedly
rich chemistry. We propose a new observational classification based on the 3~mm molecular
line emission normalized by the core visual extinction (\Av). Based on this classification,
we report a clear differentiation in terms of chemical composition and of line emission
properties, which served to define three molecular core groups. The ``diffuse'' cores,
\Av$\la$15, show poor chemistry with mainly simple species (e.g. \cs\ and \cdh).  The
``oxo-sulfurated'' cores, \Av$\simeq$15--22, appear to be abundant in species like \so\ and
\sod, but also in  HCO, which seem to disappear at higher densities. Finally,  the
``deuterated'' cores, \Av$\gsim$22, show typical evolved chemistry prior to the onset of
the star formation process, with nitrogenated and deuterated species, as well as carbon
chain molecules. Based on these categories, one of the ``diffuse'' cores (Core~47) has the
spectral line properties of the ``oxo-sulfurated'' ones, which suggests that it is a
possible failed core.}

   \keywords{	ISM: individual objects: Pipe Nebula -- 
		ISM: lines and bands --
		ISM -- stars: formation}

   \maketitle
%
%________________________________________________________________

%%%%%%%%%%%%%%%%%%%%%%%%%%%%%%%%%%%%%%%%%%%%
   \begin{figure*}[ht]
   \includegraphics[width=\textwidth,angle=0]{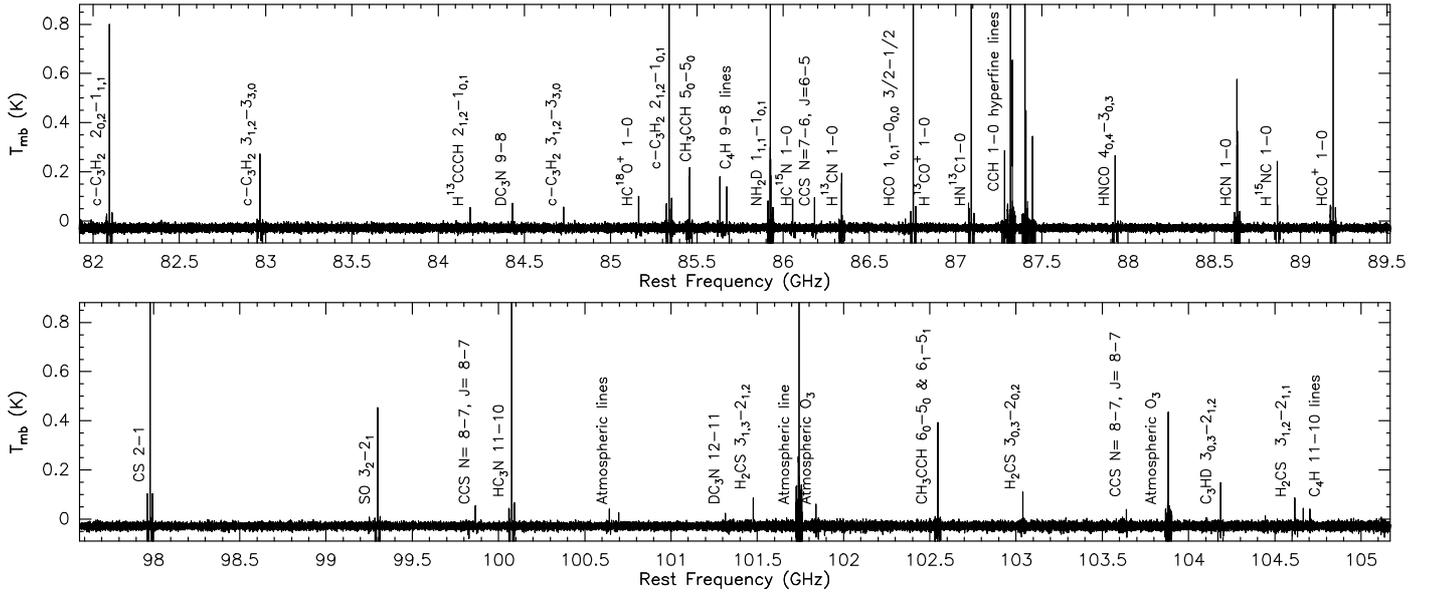}

      \caption{IRAM~30-m EMIR+FTS full bandwidth spectrum toward Core~12.  The most
      important detected molecular transitions are labeled within the plot. The upper and
      lower panels show the $\sim$7.6~GHz lower and upper sidebands, respectively.  The
      noise rarely exceeds 10~mK. The negative emission are the twin negative counterparts
      of the positive emission due to the frequency-switching observing mode.}

       \label{fig_c12}
   \end{figure*}
%%%%%%%%%%%%%%%%%%%%%%%%%%%%%%%%%%%%%%%%%%%%

\section{Introduction}

A new generation of sensitive receivers and wideband backends allows to study in detail the
chemistry of faint starless cores. Several surveys have been performed toward them
reporting a rich but relatively simple chemistry: essentially carbon chemistry with
significant sulfur and nitrogen bearing molecules, followed by later deuteration which can
be used as a chemical clock (e.g., 
\citealp{turner94,turner00,hirota06,tafalla06,bergin07}). Recently, from the theoretical
side, several papers have tried to model the starless core chemistry self-consistently
\citep{aikawa01,garrod05,keto08}.

The \pipe\ is a nearby (145~pc: \citealp{alves07}) cloud  that harbors more than one
hundred of low mass ($\sim$1~M$_{\odot}$) starless cores, most of them gravitational
unbound but confined by the thermal/magnetic pressure  of the whole cloud
\citep{alves08,lada08,franco10}. The \pipe\ differs from the other nearest dark cloud 
complexes such as Taurus  or $\rho$~Ophiuchus because it has a very small star formation
efficiency \citep{onishi99,forbrich09,roman10,roman11}.   Thus, the \pipe\ is an ideal
target to study the physical and chemical conditions in a pristine environment prior to the
onset of the star formation process, as the recent numerous studies have shown (e.g.,
\citealp{brooke07,muench07,rathborne08}). \citet{frau10} present the first results of an
extensive continuum and molecular line study on a subset of a selected sample of cores
distributed in the different regions of the \pipe: {\it bowl}, {\it stem}, and B59.  The
cores are in general less dense and less chemically evolved than starless cores in other
star forming regions studied (e.g. \citealp{crapsi05}).  We find very different
morphologies and densities, and  no clear correlation of the chemical evolutionary stage of
the cores with their  location in the cloud.  The \pipe\ starless cores have shown to be
more heterogeneous than expected.

In this work, we present a wide ($\sim$15~GHz) unbiased chemical survey at 3~mm toward a
larger sample of \pipe\ starless cores, spanning a factor of 6 in their visual extinction
(\Av) peaks. This is a first step to characterize their varied chemistry in order to
proceed to future modeling.

%__________________________________________________________________

\section{FTS Observations and Data Reduction}

We made pointed observations toward the \pipe\ Cores~06, 08, 12, 14, 20, 33,  40, 47, 48,
56, 65, 87, 102 and 109, following \citet{rathborne08} numbering, and toward a position
with no cores. We pointed either toward the continuum emission peak \citep{frau10,frau11},
if available, or toward the \cdo\ pointing center reported by \cite{muench07}. We assumed
that the pointing centers were the densest region of the cores and, therefore, with the
richest chemistry. We used the EMIR heterodyne receiver of the IRAM 30-m telescope tuned at
the C$_2$H~(1--0) transition (87.3169~GHz). At this frequency the telescope delivers
$\theta_{HPBW}$=28\asec1, $B_{\rm eff}$=0.81 and $F_{\rm eff}$=0.95. The observations were
carried out in August 2011, being the first astrophysicists to use the FTS autocorrelator
as the spectral backend. We selected a channel resolution of 195~kHz ($\simeq$0.6~\kms at
3~mm) which provided a total bandwidth of 14.86~GHz covering the frequency ranges from
82.01 to 89.44~GHz, and from 97.69 to 105.12~GHz. We used the frequency-switching mode with
a frequency throw of 7.5~MHz. System temperatures ranged from $\sim$110 to $\sim$150~K.
Pointing was checked every 2 hours. We reduced the data using the CLASS package of the
GILDAS\footnote{available at http://www.iram.fr/IRAMFR/GILDAS} software. The baseline in
the frequency switching mode for such a large bandwidth ($\sim$3.7~GHz for each chunk) has
a complicated shape with sinusoidal-like ripples. Nevertheless, since the observed lines
are very narrow ($\la$0.5~\kms, similar to the effective spectral resolution), the
baselines can be efficiently removed if  small windows are used ($\la$20~MHz). The
resulting typical rms noise was $\simeq$8~mK at the 195~kHz spectral resolution.

%______________________________________________________________

\section{Results}

The large available bandwidth, 14.86~GHz, has allowed us to carry out an unbiased survey,
covering about a third of the observable 3~mm window. We used the {\it
Splatalogue}\footnote{http://www.splatalogue.net/} tool to identify possible  lines.  We
consider tentative detections those lines with intensities in the 3--5$\sigma$ range, and
fiducial detections those higher than 5$\sigma$. We have detected 53 transitions from a
total of 31 molecules (including isotopologues). Figure~\ref{fig_c12} shows the observed
bandwidth toward  Core~12 with the main molecular species labeled.  Most of the detected
lines were identified in this core, the one with the highest \Av\ and the brightest
molecular line emission of the sample.  However, there are few sulfur bearing molecular
lines  not detected toward Core~12 but present in other cores:  \sod~$3_{1, 3}$--$2_{0,
2}$,  \tso~$3_2$--$2_1$, and \ocs~7--6.  We report tentative detections ($\sim$4$\sigma$)
of \hocom\ (Cores~6 and 102),  l-C$_3$H (Cores~12 and 109), and \hccnc\ (Core~12).  We have
also identified in all the cores several Earth atmospheric lines, mostly  from ozone. 

The cores with the brightest detected lines are those with highest \Av\ (Cores~12, 87 and
109) due to their larger gas column densities. In order to avoid a column density bias (our
core sample spans a factor of 6 in \Av), we have normalized the intensity by dividing the
spectra by the \Av\ peak of the core. We used the values obtained by \citet{roman10} from
dust extinction maps that have an angular resolution similar to our observations. This
definition mimics molecular abundances for optically thin lines. Figure~\ref{fig_spec}
shows a selected sample of the brightest normalized lines toward all the sample, with the
cores ordered by their \Av\ peak. In this figure we have ordered the molecules in families
taking into account their atomic composition.

In general, the 3~mm transitions of the lightest species of most of the molecular families
(blue spectra in Fig.~\ref{fig_spec}) were detected in all the cores of our sample: \cdh,
\hcom, \cs, \so, and \hcn. \cthd\ can also be part of this sample, since it was detected in
all but two cores. The 3~mm main transitions of these molecules can be considered
``ubiquitous lines'' in starless cores. \hcom~1--0, \cs~2--1, and \hcn~1--0 show little
variations in normalized intensity.  These molecules have large dipole moments and high
abundances, therefore they likely have large optical depths \citep{frau11}. In addition,
the \hcom~1--0 and \hcn~1--0  transitions can be affected by absorption by low density
foreground gas \citep{girart00}. Indeed, the relative \hcn~1-0 hyperfine line intensities
of Cores~12, 40 and 87 suggest that this transition is out of LTE.  The normalized
intensities of the other three ubiquitous lines  show significant variations within the
sample. However, while \cdh~1--0 and \cthd~$2_{1,2}$--$1_{0,1}$  tend to increase with \Av,
the \so~3$_2$--2$_1$ line appears to have the largest normalized intensities in the cores
with visual extinction in the range of 15 to 22 mag.

Several molecular transitions were detected only toward cores with \Av$\gsim$15.  The
optically thin \htcom~1--0, and \hcdom~1--0, and the transition 
\hnco~$4_{0,4}$--$3_{0,3}$, show larger normalized intensities with increasing  column
densities.  The detected transitions from oxo-sulfurated molecules (\so~2$_2$--1$_1$,
\tso~3$_2$--2$_1$, \sod~$3_{1,3}$--$2_{0,2}$ and \ocs~7--6)  are mainly detected toward the
cores with the brightest \so~3$_2$--2$_1$ emission, this is, mainly in the cores with
\Av$\simeq$15--22 mag. The \hco~1--0 transition shows the same behavior. Curiously and
despite its low density (\Av=11.2~mag), Core~47 shows emission in most of the
oxo-sulfurated molecular transitions as well as in \hco~1--0.  The
\hdcs~3$_{1,3}$--2$_{1,2}$ transition appears to show a similar trend to the oxo-sulfurated
molecules, although it peaks at slightly denser cores (\Av$\simeq$20~mag) and clearly
survives at larger \Av\ values. The emission of the other two lines of this group,
\hcsp~2--1 and \tcdo~1--0, is too  weak to show a clear trend.

The number of detected molecular transitions increased significantly for the  four cores
with the highest column density (\Av$\ga 22$ ~mag) due to either ({\it i})
excitation/column density reasons or ({\it ii}) synthesization  timescales. The \cthd\
molecule is a good example of the former molecules.  Although being ubiquitous in the
$2_{1,2}$--$1_{0,1}$ transition, the $3_{1,2}$--$3_{0,3}$ one is only detected at these
column densities.  The rarer isotopologic counterparts of the \hcn\ and \cthd~1--0
ubiquitous lines (\htcn, \hcqn,  and \tcthd) are detected only in these four cores. This is
also the case for \hntc\ and \hqnc\ in the 1--0 transition, which suggests that the
\hnc~1--0 is likely to be an ubiquitous line as well. Most of the transitions detected in
these four cores show larger normalized  intensities  with increasing \Av\ (e.g.
\hctn~11-10).  The carbon-chain molecular transitions (\cqh~9--8 and  11--10, and
\chtcdh~5$_n$--4$_n$ and 6$_n$--5$_n$) are the exception, showing  little variations  in
normalized intensity.  Several transitions of three deuterated forms of abundant species
were also detected:  \cthdt~$3_{0,3}$--$2_{1,2}$, \nhdd~1-1, and \dctn\ in the 9--8 and
12-11  (see Fig.~\ref{fig_spec}). Only the first transition is detected in the four cores.

%______________________________________________________________

\section{Discussion and Conclusions\label{disc}}

\begin{landscape}
   \begin{figure}
   \includegraphics[height=15cm,angle=0]{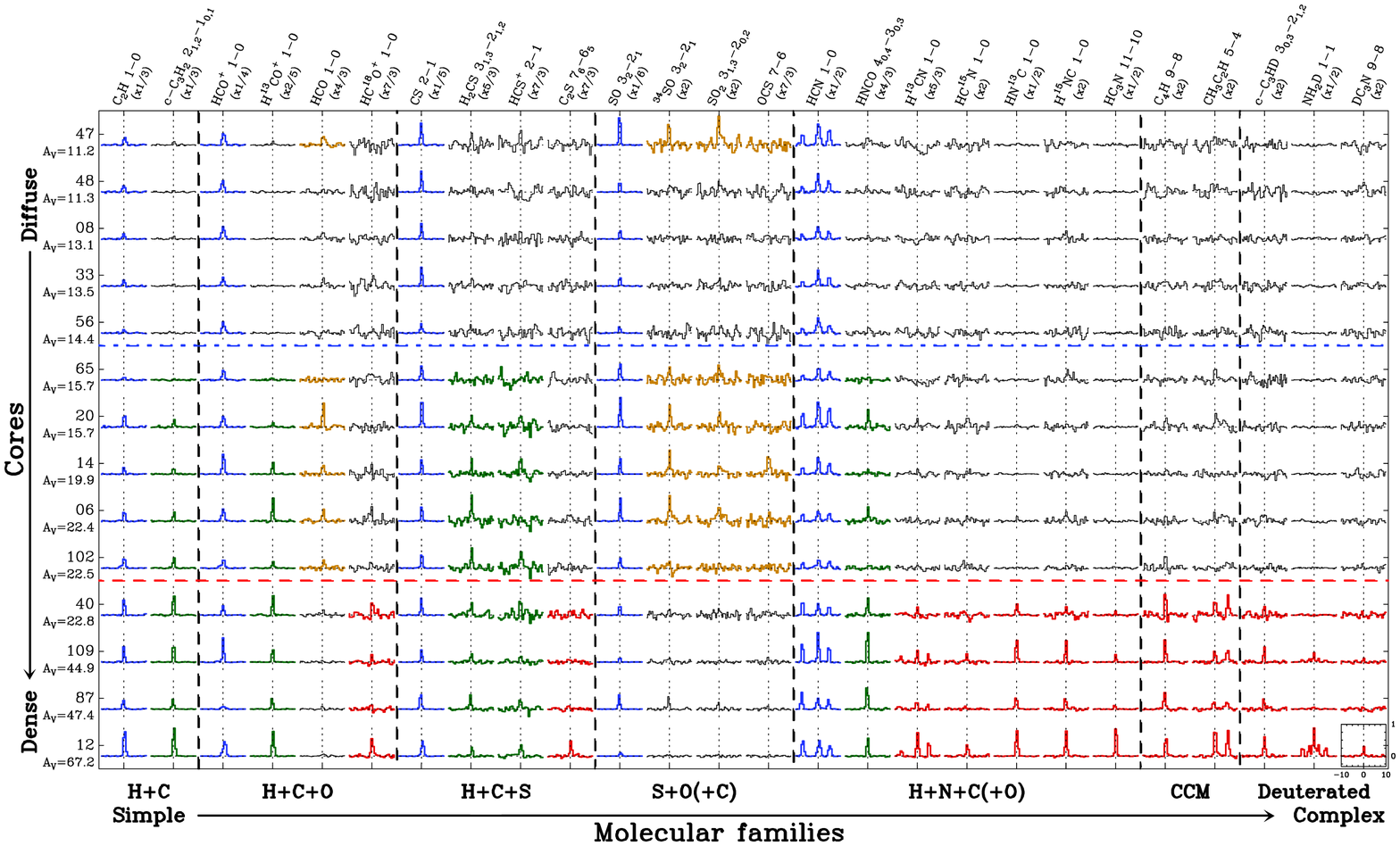}

      \caption{ Selected normalized molecular transitions toward the observed cores. The
      scale is shown in the bottom right spectrum. The normalized intensity axis ranges
      from -0.33 to 1, while the velocity axis spans 20~\kms\ centered at the \vlsr. {\it
      Rows}:  individual cores, labeled on the left-hand side of the figure, ordered by its
      \Av\ peak. {\it Columns}: molecular transition, ordered by molecular families,
      labeled on the top of the figure. The spectra have been divided by $[ \Av/100 \, {\rm
      mag}]$ to mimic the abundance, where the \Av\ value is that at the respective core
      center \citep{roman10} given below the core name. Each molecular transition has been
      multiplied by a factor, given below its name, to fit in a common scale. {\it Colors}:
      used to highlight the distinctive emission of the different core groups.
      \textcolor{blue}{blue}: ubiquitous lines, \textcolor{ForestGreen}{green}:
      dense-medium molecular transitions, \textcolor{BurntOrange}{orange}: molecular
      transitions typical in  oxo-sulfurated cores (see Sect.~\ref{disc}),
      \textcolor{BrickRed}{red}: molecular transitions typical in  deuterated cores, and
      {\bf black}: mostly undetected species.\label{fig_spec}}

   \end{figure}
\end{landscape}

The chemistry detected toward the sample of fourteen starless cores is unexpectedly rich 
taking into account their low temperatures (10--15 K: \citealp{rathborne08}) and visual
extinctions. The apparent correlation within the sample of the 3~mm molecular transition
normalized intensities to visual extinction allow us to propose an observational
classification (see Fig.~\ref{fig_spec}). We define three groups of starless cores, which
are probably related with their dynamical age: {\it ``diffuse''},  {\it
``oxo-sulfurated''}, and {\it ``deuterated''} cores. This classification can be useful in
future wide band 3~mm observations of molecular clouds.

{\it -- ``Diffuse'' cores}: a set of cores with small column densities (\Av$\lsim$15~mag
$\sim$ $N_{\rm H_2}$$\lsim$1.2$\times$10$^{22}$~cm$^{-2}$) lies above the blue dot-dashed
horizontal line in Fig.~\ref{fig_spec}.  Their spectra is rather poor, showing only
significant normalized intensity in the  transitions of the main isotopologues of abundant
species like \cdh,  HCN (and likely HNC), \hcom\ and SO.  Such a simple observational
chemistry suggests that these are very young starless cores, or even transient clumps on
which essentially the cloud chemistry is better detected due to density enhancements.
Core~47 is a clear exception and it is discussed later in the text.

{\it -- ``Oxo-sulfurated'' cores}: a group of denser cores  (\Av$\simeq$15--22~mag  $\sim
N_{\rm H_2} \simeq$1.2$\times$10$^{22}$--1.7$\times$10$^{22}$~cm$^{-2}$) that show richer
chemistry but not yet significant deuteration to be observed.  In Fig.~\ref{fig_spec} this
group  lie between the  blue dot-dashed and red dashed horizontal lines. All the
transitions detected in the {\it ``diffuse''} cores are also present.  The \so~3$_2$--2$_1$
transition is the main signpost as it is very  bright. Many other oxo-sulfurated molecules
(\tso, \sod, and \ocs), as well as \hco, exhibit a same trend, but they are not detected at
higher densities. This suggests an increase of these chemically related species in the
gas-phase in this \Av\ range, followed by a later depletion/destruction as density
increases. These cores might be in-the-making cores, which have developed a richer
chemistry  and piled up more material, probably in a stage near to the onset of collapse
\citep{ruffle99}.   Core~102 is an exception in this group and is discussed later in the
text.

{\it -- ``Deuterated'' cores}: the densest cores of the sample (\Av$\gsim$22~mag $\sim$
$N_{\rm H_2}$$\gsim$1.7$\times$10$^{22}$~cm$^{-2}$), shown below the red dashed horizontal
line in Fig.~\ref{fig_spec}. Core~12, the densest one, sets the upper limit at \Av=67.2~mag
($N_{\rm H_2}$$\simeq$5.3$\times$10$^{22}$~cm$^{-2}$). These cores are generally bright in
the transitions typical of the other two groups. The oxo-sulfurated molecules are the
exception, hardly present, probably depleted/destructed at the densities reached. The main
signpost are the emission, only present in this group, in rare isotopologues of the
nitrogenated ubiquitous lines (\htcn, \hcqn, \hntc, and \hqnc), deuterated forms of
abundant species (\cthdt, \nhdd, and \dctn), and carbon-chain molecules (\cqh\ and
\chtcdh).  These cores might be stable starless cores with a life-time long enough to
achieve the densities needed to synthesize efficiently carbon chains and deuterated species
\citep{roberts00,gwenlan00}. 

As already said, Core~47 does not match the chemical properties of the diffuse cores. It
shows a chemistry matching the oxo-sulfurated group, which proved to be very sensitive to
density. This suggests that it might be a failed core which developed a rich chemistry and
is now merging back into the cloud. This scenario can increase the oxo-sulfurated chemistry
detected \citep{garrod05}.  Core~47 is located close to Core~48 in the only \pipe\ region
with superalfv\'enic turbulence, as shown by optical polarization observations
\citep{franco10}. Therefore, it is possible that an external source of turbulence is
disrupting the medium in this area and dispersing the cores.

On the other hand, Core~102, in the oxo-sulfurated group, shows a chemistry similar to that
of the diffuse cores. Similarly, Core~87, among the deuterated cores, shows features of the
oxo-sulfurated group. This suggests that these cores might have piled up material so
quickly that a more complex chemistry had no time to be synthesized. Both cores lie in the
same N-S oriented high-density structure \citep{roman10} where \citet{franco10} report a
N-S magnetic field. The fast evolution might have been driven by magnetic fields with the
surrounding mass collapsing in this direction.

The FTS chemical survey toward the starless cores of the \pipe\ showed a chemistry much
more rich than expected for a cloud giving birth to low-mass stars at very low efficiency.
A good interpretation of the results demands chemical modeling to investigate a possible
evolutive track, which will be the purpose of a forthcoming study.

%______________________________________________________________

\begin{acknowledgements}

PF is partially supported by MICINN fellowship FPU (Spain). PF, JMG and MTB are supported
by MICINN grant AYA2008-06189-C03 (Spain). PF, JMG and MTB are also supported by AGAUR
grant 2009SGR1172 (Catalonia). We thank Carlos Rom\'an-Z\'u\~niga for gently sharing the
\Av\ maps. The authors want to acknowledge all the IRAM 30-m staff for their hospitality
during the observing runs, the operators, and the AoDs for their active support.  We thank
the anonymous referee for useful comments. This research has made use of NASA's
Astrophysics Data System.

\end{acknowledgements}

%______________________________________________________________

%%%%%%%%%%%%%%%%%%%%%%%%%%%%%%%%%%%%%%%%%%%%
%%%%%%%%%%%%%%%%%%%%%%%%%%%%%%%%%%%%%%%%%%%%

\end{document}